\documentstyle[graphics]{jaa}

\begin{document}
\title[Position angle changes of inner-jets of blazars]
      {Position angle changes of inner-jets in a sample of blazars \\}
\author[Ligong Mi \& Xiang Liu]%
       {Ligong Mi$^{1,3,*}$, Xiang Liu$^{1,2}$\\
        $^1$ Xinjiang Astronomical Observatory, CAS, 150 Science 1-Street, Urumqi 830011, PR China\\
        $^2$ Key Laboratory of Radio Astronomy, CAS, Nanjing 210008, PR China\\
        $^3$ University of Chinese Academy of Sciences, Beijing 100049, PR China\\
        $^*$ email: miligong@xao.ac.cn\\}
\date{}
\maketitle
\label{firstpage}
\begin{abstract}We have carried out the Gaussian model-fitting to
the 15~GHz VLBA cores for a sample of blazars from the MOJAVE
database, analyzed the correlations in the model-fitted parameters
and studied the variability properties for different group of
sources. We find that the {\it Fermi} LAT-detected blazars have on
average higher position angle changes of cores than the LAT
non-detected blazars, and that the LAT-detected ones largely associate
with more variable cores in flux density.
\end{abstract}

\begin{keywords}
Active galactic nuclei: blazars -- radio continuum: variability --
$\gamma$-ray: {\it Fermi}-LAT
\end{keywords}

\begin{centering}
\section{ Model-fit and statistical analysis}
\end{centering}
\label{sec:intro}

\noindent We define a sample of 104 blazars that have been
monitored for more than 10 years with the VLBA at 15~GHz till end of
2011 and have at least 15 epoch data sets with good time
coverage, from the MOJAVE database (Lister et al. 2009).
The 104 blazars all show prominent core-jet features in the 15 GHz
VLBA images. We have model-fitted to the `core' component of the
natural-weighted 15 GHz VLBA images for each blazar, with a
two-dimensional elliptical Gaussian model in the AIPS task `JMFIT',
to obtain the position angle (PA) of the major axis of the elliptic
Gaussian component, the integrated flux-density (IF) of the
Gaussian component, and the de-convoluted major and minor axis
scales of the Gaussian component. We consider that the core
component of 15 GHz VLBA image is an `inner-jet' rather than a
true core, and the inner-jet could be modelled with an elliptical
Gaussian component with its major axis along the
inner-jet orientation or the inner-jet ridge-line on average, thus
reflecting the inner-jet position angle. As demonstrated by Liu et
al. (2012), the model-fitted position angles of cores are not
correlated at all with the position angles of the VLBA restoring
beams.

\begin{figure*}
\centering
\begin{tabular}{c}
\hfill\includegraphics{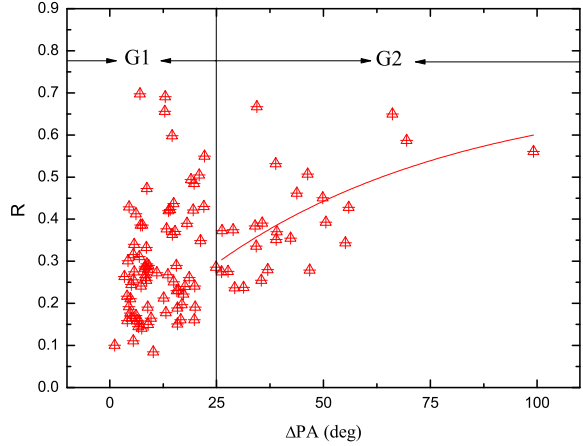}%
\end{tabular}
\caption{Relationship between the average of the minor/major axis ratio, R, and
the position angle changes of inner-jet, $\Delta$PA. A positive
correlation between R and $\Delta$PA appears for
$\Delta$PA$>$$25^{\circ}$ (G2), and there is no
correlation for $\Delta$PA$<$$25^{\circ}$ (G1).} \label{fig1}
\end{figure*}

We define $\Delta$PA as the difference between the median and
minimum of the PA, and use it as a measure of the position angle
changes of inner-jet. It is expected that the Gaussian
model-fitted PA of inner jet could be arbitrary when the
minor/major ratio of the Gaussian component is close to unity, but
this error in PA cannot be properly estimated by the model-fit
task itself (see Condon 1997). To test this, we define an average
of the minor/major ratio for each blazar core, R, and find a positive
correlation between $\Delta$PA and R for $\Delta$PA$>$$25^{\circ}$, which supports that the
error in the model-fitted PA is proportional to the
minor/major ratio of core. In the regime of smaller $\Delta$PA
and/or smaller R, this effect is not significant, e.g. there
seems to be no significant correlation between $\Delta$PA and R for
$\Delta$PA$<$$25^{\circ}$. For simplicity, we divide the 104
sources into group 1 for $\Delta$PA$<$$25^{\circ}$ and group 2 for
$\Delta$PA$>$$25^{\circ}$ (G1 and G2, respectively, as shown in Fig.~\ref{fig1}),
and we only consider the group 1 in the following, since the group 2
could have relatively larger errors in inner-jet position angles.
The G1 contains 57 Quasars and 21 BL Lac objects, of which 58
blazars are {\it Fermi} LAT-detected and 20 blazars are the LAT
non-detected. To analyze possible correlation between the IF and
the PA for the G1, we define a significant linear correlation as
an absolute value of linear correlation coefficient$>$0.40 at
confidence of $>$95\%. About 25\% (20 out of 78) blazars show
significant linear correlations between IF and PA. The details are
shown in Table 1. The columns are (1) source type; (2) source count;
(3),(4) number (fraction) of sources having positive, negative
correlation between IF and PA, and a `+' or `-' sign indicates a
positive or negative correlation respectively. Non-ballistic
counter-clockwise (or clockwise) helical jet models mainly basing
on the geometric beaming effect could respectively explain the
positive correlation (or the negative correlation) between the IF and
the PA of inner-jets of the blazars (Liu et al. 2012).

\begin{table}[htbp]
\caption[]{\label {tab:test}Statistics of correlations of IF vs PA
in G1} \centering
\begin{tabular}{cccc}
 \noalign{\smallskip}
 \hline
  1&2 &3 &4   \\
\hline
Class & N & Positive corr.& Negative corr.\\
\hline
Blazars&78&+12(15\%)&-8(10\%)\\
\hline
Quasars&57&+8(14\%)&-8(14\%)\\
BL Lacs&21&+4(19\%)&-0(0\%)\\
\hline
\end{tabular}
\end{table}

\begin{figure*}
\centering
\begin{tabular}{c}
\hfill\includegraphics{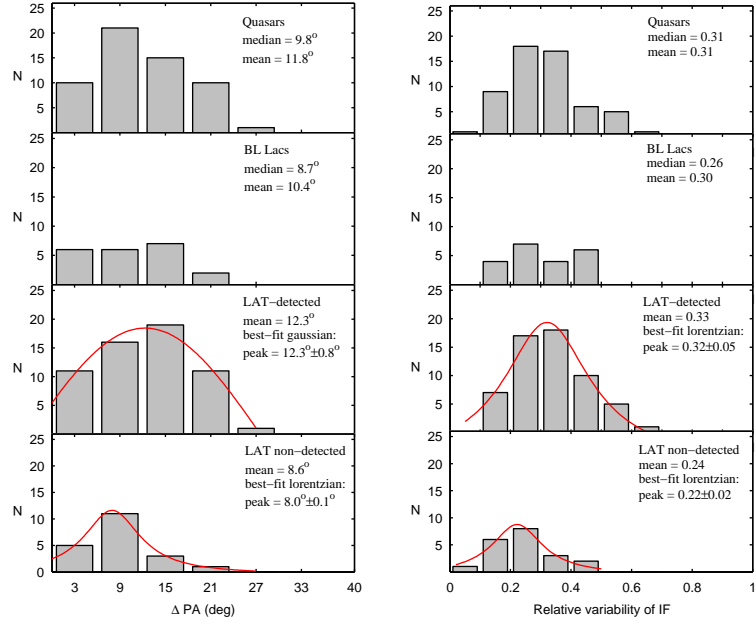}%
\end{tabular}
\caption{Left: {\it Distributions} of $\Delta$PA for 57 quasars,
21 BL Lacs, 58 LAT-detected and 20 LAT non-detected respectively
from top to bottom panel. Right: {\it Distributions} of the
relative variability (from zero to 100\%) of integrated flux density (IF)
for 57 quasars, 21 BL Lacs, 58 LAT-detected and 20 LAT non-detected
respectively from top to bottom panel.}
\label{fig2}
\end{figure*}

We also study the long-term variability of the parameters: PA and
IF. The distributions of the $\Delta$PA for quasars, BL Lacs, the
LAT-detected and the LAT non-detected are shown in Fig.~\ref{fig2}
(left). Gaussian or Lorentzian function fittings are used for
the histogram distribution in order to obtain the peak value. The
Kolmogorov-Smirnov (K-S) tests for 57 quasars and 21 BL Lacs
suggest that the two distributions are not significantly
different; there appears to be a peak at higher $\Delta$PA in the
BL Lacs than in the quasars, but not significantly. The K-S tests
for 58 LAT-detected and 20 LAT non-detected indicate a low
probability (p $=$ 0.012) for these two samples being drawn from
same parent population. It displays that the source distribution
peaks at $\Delta$PA of $12.3^{\circ}\pm0.8^{\circ}$ in the
LAT-detected is higher than the peak of $\Delta$PA
($8.0^{\circ}\pm0.1^{\circ}$) in the LAT non-detected, suggesting
that the $\gamma$-ray blazars mostly have a larger $\Delta$PA of
inner-jet than the non-$\gamma$-ray blazars.

Source counts versus the `relative variability of IF' (defined as
long-term rms-variation divided by the mean of IF) are displayed in
Fig.~\ref{fig2} (right). The K-S tests for 57 quasars and 21 BL
Lacs suggest that the two distributions are not significantly
different. The K-S tests for 58 LAT-detected and 20 LAT
non-detected indicate a low probability (p $=$ 0.005) for these
two samples being drawn from same parent population. It displays
that the source distribution peaks at the relative variability of
$0.32\pm0.05$ in the LAT-detected is higher than the peak of
$0.22\pm0.02$ in the relative variability of the LAT non-detected,
suggesting that the $\gamma$-ray blazars mostly have a larger
relative variability of inner-jet than the non-$\gamma$-ray blazars.\\

\section*{\centering Acknowledgements}
We thank the anonymous reviewer for helpful comments and
suggestions on the manuscript. This research has made use of data
from the MOJAVE database that is maintained by the MOJAVE team
(Lister et al., 2009, AJ, 137, 3718). The work is supported by the
National Natural Science Foundation of China (Grant No.11273050)
and the 973 Program of China (2009CB824800).

\begin{centering}

\end{centering}
\end{document}